\documentstyle[12pt]{article}
\def\jp{J/\psi}
\begin{document}

\begin{center}
{\Large \bf Tests of Flavor Symmetry in $J/\psi$ Decays}\\[1.5cm]
\end{center}
\begin{center}
{{\large \bf G. L\'opez Castro$^1$, J. L. Lucio M.$^2$ and J. 
Pestieau$^3$}\\[1cm]
$^1$ {\it Depto. de F\'\i sica, Cinvestav del IPN, M\'exico DF. MEXICO} \\
$^2$ {\it Instituto de F\'\i sica, Univ. de Guanajuato, Le\'on-Gto. MEXICO}\\
$^3$ {\it Inst. de Phys. Th\'eorique, UCL, Louvain-la-Neuve. BELGIUM}}
\end{center}
\noindent
\begin{center}
\begin{small}
\begin{minipage}[t]{12cm}
{\bf Abstract.} We use SU(3) flavor symmetry to analyze the $PP',\ VP$ 
and baryon-antibaryon decays of $J/\psi$. Both, the SU(3)-invariant and 
-violating contributions are considered.  Particular attention is paid to 
the interference of the electromagnetic and strong amplitudes.
\end{minipage}
\end{small}
\end{center}
\setlength{\baselineskip}{1\baselineskip}

\begin{center} 
\bf INTRODUCTION
\end{center}
\hspace*{.5cm}The $J/\psi$ was discovered simultaneously in $e^+e^-$ and 
$p\bar{p}$ 
collisions in 1974 (1). Twenty years later, more than one hundred of its 
exclusive decay modes are reported by the Particle Data Group (2). The 
measurements of these decays  serve several purposes (3): the study of 
light hadron spectroscopy, the $q\bar{q}$ content of mesons, the 
electromagnetic form factors at the $J/\psi$ mass, the determination of 
$\alpha_s$, the pattern of flavor symmetry breaking, the search for 
exotics ($gg$, hybrids and $4q$ states), among others.\\
\hspace*{.5cm} The $\jp$ is only 118 MeV above the $\eta_c(1S)$, which is 
the lowest 
lying state of the charmonium family. This means that all of its decay 
modes, other than $\jp \rightarrow  \eta_c \gamma$, are suppressed by 
the so-called OZI rule (4).  This suppression occurs because 
 the final states containing only hadrons are reached through the  
(a) electromagnetic (1 virtual photon) and, (b) strong 
( 3 gluons) annihilation of $J/\psi$. \\
\hspace*{.5cm}  The $J/\psi$ system offers a unique opportunity to study the 
simultaneous effects of 
electromagnetic and strong interactions. On the one hand, since
$\alpha_s^3(M_{J/\psi}) \approx \alpha_{em}$, this suggest that  
electromagnetic and strong annihilation amplitudes for $J/\psi$ hadronic 
decays could have  
similar strenghts. Moreover, the $J/\psi$ has an special status among the quarkonium vector 
states because it is three times heavier than the $\phi (s\bar{s})$,
where non-perturbative QCD is expected to dominate the OZI suppressed decays and, on the other 
hand, it is three times lighter than the $\Upsilon (b\bar{b})$, where 
perturbative QCD  plays 
the main role to produce hadrons.\\
\hspace*{.5cm}   In this work we are concerned with tests of SU(3) flavor 
symmetry\footnote{Hereafter SU(3) 
refers to the flavor symmetry of light hadrons.} in two-body hadronic 
decays of $J/\psi$\footnote{Related analysis but with a different 
emphasis, can be found in Refs. (6-8)} (5). 
The present analysis consider $J/\psi$ decays into the $PP',\ VP$ and $B
\overline{B}$ channels ($P,\ V$ and $B$ for a pseudoscalar, vector and spin-1/2 baryon, 
respectively).  Our main focus will be on the relative size of the 
$1\gamma$ and 
$3g$ contributions and on the relative phase between them (6). 
This would be 
important in order to isolate the relevant electromagnetic form factors 
at the $J/\psi$ mass. \\
 \hspace*{.5cm}  Our starting point is to realize that the $J/\psi$ is a 
singlet under SU(3). 
The main advantage of doing an SU(3) analysis of the $J/\psi$ hadronic 
decays relies 
on the fact that both, the light final state hadrons and the symmetry 
breaking interactions,  have well defined 
transformation properties under SU(3). For definiteness let us introduce two important 
ingredients of the analysis:\\
\hspace*{.3cm}$(a)$ The relevant decay amplitudes must include the 
effects of SU(3) breaking. At 
the fundamental level, the sources of this flavor symmetry breaking are the quark-mass 
differences and the electromagnetic interactions\footnote{ For simplicity we neglect the isospin 
breaking in the quark mass sector ($m_u=m_d=m$).} namely:
\begin{equation}
{\cal H}_m = \frac{1}{2} \overline{q} \{ a\lambda_0 + b\lambda_8\}q
\end{equation}
and
\begin{equation}
{\cal H}_{em} = \frac{e}{2} A^{\mu}\overline{q} \gamma_{\mu} \{ \lambda_3 + 
\frac{\lambda_8}{\sqrt{3}}\} q,
\end{equation}
where $q^T = (u,d,s),\ A^{\mu}$ is the electromagnetic four-potential, $\lambda_{3,8}$ are 
Gell-Mann matrices, $\lambda_0=\sqrt{2/3}I$ and $a=\sqrt{2/3}(m_s + 2m),\ 
b=-2(m_s-m)/\sqrt{3}$. \\
\hspace*{.3cm}$(b)$ Following Ref. (7), it is useful to introduce a 
generalized charge conjugation operation 
${\cal C}$. According to ${\cal C}$, $J/\psi$ decays into a pair of hadrons that belong to the 
same representation of SU(3) are forbidden in the SU(3) limit. Thus, the  $J/\psi 
\rightarrow PP'$ decays are forbidden in the SU(3) limit, while $J/\psi \rightarrow VP,\ 
B\overline{B}$ are allowed.

\begin{center}
\bf TWO-PSEUDOSCALAR CHANNEL
\end{center}
\hspace*{.5cm}The measured decay modes in this channel which are 
reported by the Particle Data Group (2) 
are: $\pi^+ \pi^-,\ K^+K^-$ and $K_L K_S$. In order to use flavor SU(3), 
one introduces an 
octet of pseudoscalar fields $P^a,\ a=1,\ \cdots, 8$. The effective Lagrangian used to 
describe these SU(3)-violating decays is given by (7):
\begin{equation}
{\cal L}= f_{abc} \psi^{\mu} P^a\overline{\partial_{\mu}}P^b [g_M 
\delta^{c8} + e^{i\phi} g_E(\delta^{c3}+\frac{\delta^{c8}}{\sqrt{3}} )]
\end{equation}
where $f_{abc}$ are the SU(3) structure constants, $\psi^{\mu}$ describes the $J/\psi$ vector 
field and $g_M,\ g_E$ are the coupling constants coming from quark-mass difference and 
electromagnetism.\\
\hspace*{.5cm}   A fit to the experimental data for the $\pi^+\pi^-,\ K^+ 
K^-$ and $K_L K_S$ decay modes (2), gives:
\begin{eqnarray}
g_M &=& (8.5 \pm 0.6) \times 10^{-4}\nonumber \\
g_E &=& (7.8 \pm 0.7) \times 10^{-4} \\
\phi &=& (90 \pm 10)^0. \nonumber
\end{eqnarray}
   Thus, the electromagnetic and quark-mass difference contributions have similar strenghts as 
expected from the naive counting of couplings ($\alpha_s^3 \approx \alpha_{em}$). Observe also 
that there is not interference of both contributions in the decay rate ($\phi \approx \pi/2$).

\begin{center}
\bf VECTOR-PSEUDOSCALAR CHANNEL
\end{center}
\hspace*{.5cm}Ten exclusive decay modes have been reported in this  
channel, namely (2): $\rho^0 \pi^0, 
K^{*+}K^-,\ K^{*0}K^0,\ \omega \eta,\ \omega \eta',\ \phi \eta,\ \phi \eta',\ \rho^0 \eta,\ 
\rho^0 \eta'$ and $\omega \pi^0$. The corresponding Lorentz invariant amplitude for this decay 
channel is given by:
\begin{equation}
{\cal M} = g_{VP} \epsilon_{\mu\nu\alpha\beta} \varepsilon^{\mu} \eta^{\nu} q^{\alpha} k^{\beta},
\end{equation} 
where $g_{VP}$ is the coupling constant, $\varepsilon^{\mu} (\eta^{\mu})$ is the $J/\psi (V)$ 
polarization four-vector and $q(k)$ the corresponding four-momentum.\\
\hspace*{.5cm}  In order to provide a description based on SU(3) one 
needs to introduce an octet ($O^a,\ 
a=1,\cdots, 8$) and a singlet ($S$) of vector and pseudoscalar fields. As 
usual, we define the following physical states:
\begin{eqnarray}
\eta &=& P_8 \cos \theta_P - P_0 \sin \theta_P \nonumber \\
\eta' &=& P_8 \sin \theta_P + p_0 \cos \theta_P \nonumber \\
\omega &=& V_8 \sin \theta_V + V_0 \cos \theta_V \\
\phi &=& V_8 \cos \theta_V - V_0 \sin \theta_V. \nonumber
\end{eqnarray}
where $\theta_P = -20^0$ and $\theta_V= \tan^{-1} (1/\sqrt{2})$ (2) denote 
the octet-singlet mixing angles.\\
\hspace*{.5cm}   Since the $VP$ channel is allowed by ${\cal C}$ both, the 
SU(3)-invariant and -violating 
contributions are present. Following Haber and Perrier (7), 
the SU(3) structure of the 
interaction Lagrangian can be written as follows:
\begin{eqnarray}
{\cal L} = \psi  &\{ &  {\bf g_8} \delta^{ab} O_1^a O_2^b + {\bf g_1} S_1 S_2 \nonumber \\
    &+& \left[ {\bf g_{M,88}} d_{ab8} + {\bf g_{E,88}} \left( d_{ab3} + 
\frac{d_{ab8}}{\sqrt{3}} \right) \right] O_1^a O_2^b \nonumber \\
 &+& \sqrt{\frac{2}{3}} \left[ {\bf g_{M,81}}O^8_1 + {\bf g_{E,81}}(O^3_1 
+ \frac{O_1^8}{\sqrt{3}}) 
\right] S_2 \\
 &+& ({\bf 81} \rightarrow {\bf 18}, 1 \leftrightarrow 2\ {\rm in \ the \ 
previous \ term}) \} \nonumber 
\end{eqnarray} 
where the subindex $E(M),ij$ in the coupling constants stands for 
electromagnetic (mass) origin and the subindex $1(2)$ 
in $O,\ S$ is for vector (pseudoscalar) states. $d_{abc}$ are the 
symmetric constants of SU(3).
  Thus, the decay amplitudes will depend on 8 free coupling 
constants as well as on the 
electromagnetic-strong relative phase $\phi$\footnote{ Observe that 
the relative phase between $g_1$ 
and $g_8$ is zero because both arise from strong interactions}.\\
\hspace*{.5cm}   If we assume nonet symmetry, we are left with only 4 free 
parameters because in this limit: \begin{eqnarray}
g &\equiv& g_8 = g_1 \nonumber \\
g_M &\equiv& g_{M,88} = g_{M,81} = g_{M,18} \\
g_E &\equiv& g_{E,88} = g_{E,81} = g_{E,18}. \nonumber
\end{eqnarray}
   Thus, in order to quantify nonet symmetry breaking in these decays we 
introduce the ratios: \begin{equation}
r_M = \frac{g_{M,81}}{g_{M,88}},\ \ \ r'_M = \frac{g_{M,18}}{g_{M,88}}
\end{equation}
and similar expressions for $r_E$ and $r'_E$.\\
\hspace*{.5cm}  A fit to the experimental data of Ref. (2), leads to:
\begin{eqnarray}
g_8 &=& (1.84 \pm 0.06) \times 10^{-3}\ {\rm GeV}^{-1} \nonumber \\
g_1 &=& (0.98 \pm 0.05) \times 10^{-3}\ {\rm GeV}^{-1} \nonumber \\
g_{M,88} &=& (3.84 \pm 1.57) \times 10^{-4}\ {\rm GeV}^{-1} \nonumber \\
g_{E,88} &=& (5.46 \pm 0.56) \times 10^{-4}\ {\rm GeV}^{-1} \nonumber \\
\phi & = & (72 \pm 17)^0 \\
r_M &=& 0.48 \pm 0.28 \nonumber \\
r'_M &=& 0.47 \pm 0.33 \nonumber \\
r_E &=& 1.23 \pm 0.16 \nonumber \\
r'_E &=& 1.36 \pm 0.24 \nonumber 
\end{eqnarray}
\hspace*{.5cm}  With the above results we can predict the isospin-violating 
double OZI-suppressed decay 
$BR(J/\psi \rightarrow \phi \pi^0) < 6.7 \times 10^{-6}$ which is consistent with the 
experimental upper limit ($< 6.8 \times 10^{-6}$) reported in Ref. (2).\\
\hspace*{.5cm}  The fit to the experimental data exhibits an interesting 
pattern. First, as for the $\jp \rightarrow PP'$ case,
 the electromagnetic $g_{E,88}$ and quark-mass $g_{M,88}$ violations of SU(3) 
have similar sizes; second, nonet symmetry breaking seems to be violated by 
almost 50 \% and, 
finally, the relative phase between electromagnetic and strong contributions is close to 
$\pi/2$. 

\begin{center}
\bf BARYON-ANTIBARYON CHANNEL
\end{center}
\hspace*{.5cm}The Particle Data Group (2) reports  measurements in five 
different decay modes ($p\bar{p},\ n\bar{n},\ 
\Lambda \overline{\Lambda},\ \Sigma \overline{\Sigma}$ and $\Xi 
\overline{\Xi}$) and an upper 
limit for $\Sigma^0 \overline{\Lambda}$. The decay amplitude for this decay 
channel can be 
written in terms of the `magnetic' ($G_M$) and the `electric' ($G_E$) 
form factors (9):
\begin{equation}
{\cal M} = \varepsilon^{\mu} \overline{u} [ G_M \gamma_{\mu} + \frac{2m(G_E - G_M)}{M^2 - 
4m^2} (q_{\bar B} - q_B)_{\mu} ] v.
\end{equation}
In terms of these form factors, the decay rate is given by:
\begin{equation}
\Gamma (J/\psi \rightarrow B\bar{B}) = \frac{G^2_M}{6\pi} \left\{ 1 + 
\frac{2m^2 x^2}{M^2} \right\} p \end{equation}
where $m$($M$) denote the mass of the baryon  ($J/\psi$) and $p$ is the 
three-momentum of $B$ 
in the $J/\psi$ rest frame. The constant $x$ is  defined as $x=G_E/G_M$.
 Some theoretical 
arguments (10) and experimental results (11) suggest that $0\leq x \leq 1$.\\
\hspace*{.5cm}    The SU(3) analysis for the couplings of $J/\psi$ to the 
octet of baryons gives the following (6): 
\begin{eqnarray}
G_{M,E} \propto \frac{\bf A}{2} \delta_{ab} &+& e^{i\phi}\left( d_{3ab} + 
\frac{d_{8ab}}{\sqrt{3}} 
\right){\bf D} + ie^{i\phi} \left( f_{3ab} + \frac{f_{8ab}}{\sqrt{3}} 
\right) {\bf F} \nonumber \\
&+& \frac{d_{8ab}}{\sqrt{3}} {\bf D'} + i\frac{f_{8ab}}{\sqrt{3}} {\bf F'}
\end{eqnarray}
where ${\bf D}$ and ${\bf F}$ (${\bf D'},\ {\bf F'}$) are the symmetric and 
antisymmetric couplings of electromagnetic 
(strong) origin, and the coupling ${\bf A}$ survives in the limit of exact 
SU(3).\\
\hspace*{.5cm}   Thus, we are left with seven free parameters: ${\bf A},\ 
{\bf D},\ {\bf F},\ {\bf D'},\ {\bf F'},\ x$ and the relative 
phase $\phi$. So, in order to perform a fit to the 5 experimental data, we 
will make a few `reasonable' assumptions:\\
 \hspace*{.2cm} $(i)$ We will set ${\bf D}=0$, because the isospin-violating 
decay $J/\psi \rightarrow \Sigma^0 
\overline{\Lambda}$ is proportional to ${\bf D}$. The current upper 
limit on the $\Sigma^0 
\overline{\Lambda}$ decay mode gives a negligible value for ${\bf D}$;\\
\hspace*{.2cm} $(ii)$ Since we do not know the value of $x$ we will fit the 
experimental data by fixing $x=0$ or $x=1$;\\
\hspace*{.2cm} $(iii)$ Finally we perform the fits by keeping fixed the 
relative phase at $\phi=0$ and $\phi=\pi/2$.\\
\hspace*{.5cm}   The results of our fits for ${\bf D}=0$ and the two 
different values of $x$ and $\phi$ are shown in Table 1 :\\

{\bf TABLE 1}. {\sf Results of the fits for the SU(3) couplings of baryons.}
\begin{table}[h]
\begin{tabular}{|c| c c c c|}
\multicolumn{5}{c}{} \\
\hline
Parameter &  $x=0$ & $x=0$ & $x=1$ & $x=1$ \\
($\times 10^{-4})$ & $ \phi=\pi/2$ & $\phi=0$ & $ 
\phi=\pi/2$ & $\phi=0$ \\ \hline
${\bf A}$ & 28.2$\pm$ 2.1 & 29.3$\pm$ 1.2 & 24.8$\pm$ 
1.9 & 25.9 $\pm$ 1.1 \\
${\bf D'}$ & --1.20 $\pm$ 2.5 & --0.08$\pm$ 1.97& --1.46$\pm$ 2.29 & --0.38 
$\pm$ 1.74 \\
${\bf F'}$ & 3.11 $\pm$ 1.88 & 1.85$\pm$ 3.70& 3.73$\pm$ 1.63 & 2.46 
$\pm$ 3.33 \\
${\bf F}$ & 6.09 $\pm$ 3.87 & 1.37$\pm$ 2.13& 5.64$\pm$ 3.40 & 1.30 $\pm$ 
1.95\\
\hline
$\chi^2$ & 8.7 $\times 10^{-3}$ & 0.20 & 1.0 $\times 10^{-2}$ & 0.23 \\
\hline
\end{tabular}
\end{table}

\hspace*{.5cm}  Although the results of the fit do not allow to draw a 
conclusion, we can observe the 
following behavior: $(a)$ the SU(3)-violating couplings ${\bf D'},\ {\bf 
F'}$ and ${\bf F}$ are 
much smaller than the SU(3)-invariant coupling ${\bf A}$, $(b)$ the 
quality of 
the fit seems better in the case $\phi =\pi/2$ although they do not 
distinguish the cases $x=0,\ 1$.\\
\hspace*{.5cm} Related to point $(b)$, we have also performed a fit (5) by 
keeping fixed the central values  but assuming a 
better accuracy (3\%) for the data. We obtain that only the model with no 
interference ($\phi=\pi/2$) survives.
\newpage
\begin{center}
\bf CONCLUSIONS
\end{center}
\hspace*{.5cm} We have performed an analysis of two-body hadronic decays 
of $J/\psi$ by using flavor SU(3). Our conclusions can be summarized as 
follows: $(a)$ the symmetry breaking contributions of electromagnetic 
and quark-mass origin turn out to have similar strenghts in  $PP',\ VP$ 
and $B\overline{B}$ channels; this is consistent with the expectations 
that the corresponding amplitudes  are of order $\alpha_{em}$ and 
$\alpha_s^3(M_{J/\psi})$, respectively. (b) The interference effects 
between the electromagnetic and strong decay amplitudes are absent in the 
decay rates analysed; this is interesting because it would help to 
isolate the relevant electromagnetic form factors at the $M_{J/\psi}$ 
energy scale. Finally, a sizable breaking of nonet symmetry shows up in 
the $VP$ decay modes.

\begin{center}
\bf ACKNOWLEDGEMENTS
\end{center}
 One of the authors (GLC) is grateful to the organizers for the very 
pleasant atmosphere at this meeting. GLC and JLLM  acknowledge 
partial finantial support from CONACyT.

 \begin{center} {\large \bf REFERENCES}
\end{center}
\small
\begin{itemize}
\item[1.] Aubert, J. J. {\em et al.}, {\it Phys. Rev. Lett.}{\bf 33}, 
1404 (1974); Augustin, J. E. {\em et al.}, {\it Phys. Rev. Lett.} {\bf 
33}, 1406 (1974).
\item[2.] Montanet, L. {\em et al.}, {\it Phys. Rev. }{\bf D50}, Part I 
(1994).
\item[3.] See for example: K\"opke, L. and Wermes, M., {\it Phys. Rep.} 
{\bf 174}, 67 (1989).
\item[4.] Okubo, S., {\it Phys. Lett.} {\bf 5}, 165 (1963); Zweig, G., 
CERN Report no. 8419/TH 412.; Iizuka, J., {\it Prog. Theor. Phys. Suppl.} 
{\bf 37-8}, 21 (1966).
\item[5.] For a more detailed analysis see: L\'opez Castro, G., Lucio M., 
J. L. and Pestieau, J. To be submitted for publication.
\item[6.] Kowalski, H. and Walsh, T. F. {\it Phys. Rev.}{\bf D14}, 852 
(1976); Rudaz, S., {\it Phys. Rev.}{\bf D14}, 298 (1976).
\item[7.] Haber, H.E. and Perrier, J., {\it Phys. Rev.}{\bf D32}, 2961 
(1985).
\item[8.] Seiden, A., Sadrozinski, H. and Haber, H. E., {\it Phys. Rev.} 
{\bf D38}, 824 (1988).
\item[9.] Bjorken, J. and Drell, S. D., {\it Relativistic Quantum 
Mechanics}, Mc Graw Hill, (1965).
\item[10.] Brodsky, S. and Lepage, G. P., {\it Phys. Rev. }{\bf D24}, 
2848 (1981); Claudson, M., Glashow, S. L. and Wise, M. B., {\it Phys. 
Rev.} {\bf D25}, 1345 (1982); Carimalo, C., {\it Int. Journal of Mod. 
Phys. } {\bf A2}, 249 (1987).
\item[11.] See for example: Eaton, M. W. {\em et al}, {\it Phys. Rev.} {\bf 
D29}, 804 (1984).
 \end{itemize} \end{document}